\documentclass{article}

\usepackage{microtype}
\usepackage{graphicx}
\usepackage{subcaption}
\usepackage{booktabs} 
\usepackage{multirow}
\usepackage{array}    
\usepackage{xcolor}
\usepackage{makecell}

\usepackage{hyperref}
\usepackage{enumitem,kantlipsum}
\usepackage{amsmath}

\newcommand{\kratio}{$\kappa_\text{ratio}$}

\usepackage[accepted]{mlsys2025}

\mlsystitlerunning{Submission and Formatting Instructions for MLSys 2025}
\begin{document}

\twocolumn[
\mlsystitle{Understanding Bottlenecks for Efficiently Serving LLM Inference With KV Offloading}



\mlsyssetsymbol{equal}{*}

\begin{mlsysauthorlist}
\mlsysauthor{William Meng}{penn, int}
\mlsysauthor{Benjamin Lee}{penn}
\mlsysauthor{Hong Wang}{int}
\end{mlsysauthorlist}

\mlsysaffiliation{penn}{University of Pennsylvania}
\mlsysaffiliation{int}{Intel}


\mlsyskeywords{Machine Learning, MLSys}


\begin{abstract}
KV cache offloading enables long-context LLM inference by storing caches in CPU DRAM, but PCIe bandwidth limitations create severe bottlenecks. In this paper, we develops an analytical framework that derives $\kappa_{\text{crit}}$, the critical cached-to-prefill token ratio where execution becomes memory-bound and show typical workloads exceed this threshold by orders of magnitude. Empirical characterization reveals 99\% of latency spent on transfers and serving offloaded requests results in GPU's consuming only 28\% of their rated TDP, motivating our proposed optimizations for hardware interconnects, model architectures, and scheduling algorithms.

\end{abstract}
]



\newcounter{mycounter}
\newcommand\showmycounter{\stepcounter{mycounter}\themycounter}

\section{Introduction}

Large language models have become essential infrastructure across domains ranging from chatbots \cite{openai} and financial analysis \cite{reddy2024docfinqalongcontextfinancialreasoning} to medical applications \cite{10.1001/jama.2023.14217}. Modern LLMs use the transformer architecture, which generates key-value (KV) caches during prefill. As models support longer contexts—hundreds of thousands to millions of tokens (LLaMA-3.1-405B \cite{grattafiori2024llama3herdmodels}, DeepSeek-R1 \cite{deepseekai2025deepseekr1incentivizingreasoningcapability}, GPT-5, Grok-4)—computational costs have increased substantially.

Tokens are often repeated between prompts in multi-turn conversations or when querying the same documents multiple times. Prefix caching \cite{kwon2023efficientmemorymanagementlarge} stores KV caches in GPU VRAM to eliminate redundant computation. However, these caches can grow to tens of gigabytes and quickly exhaust VRAM capacity when serving multiple concurrent requests.

\begin{figure}
\centering
\includegraphics[width=1\linewidth]{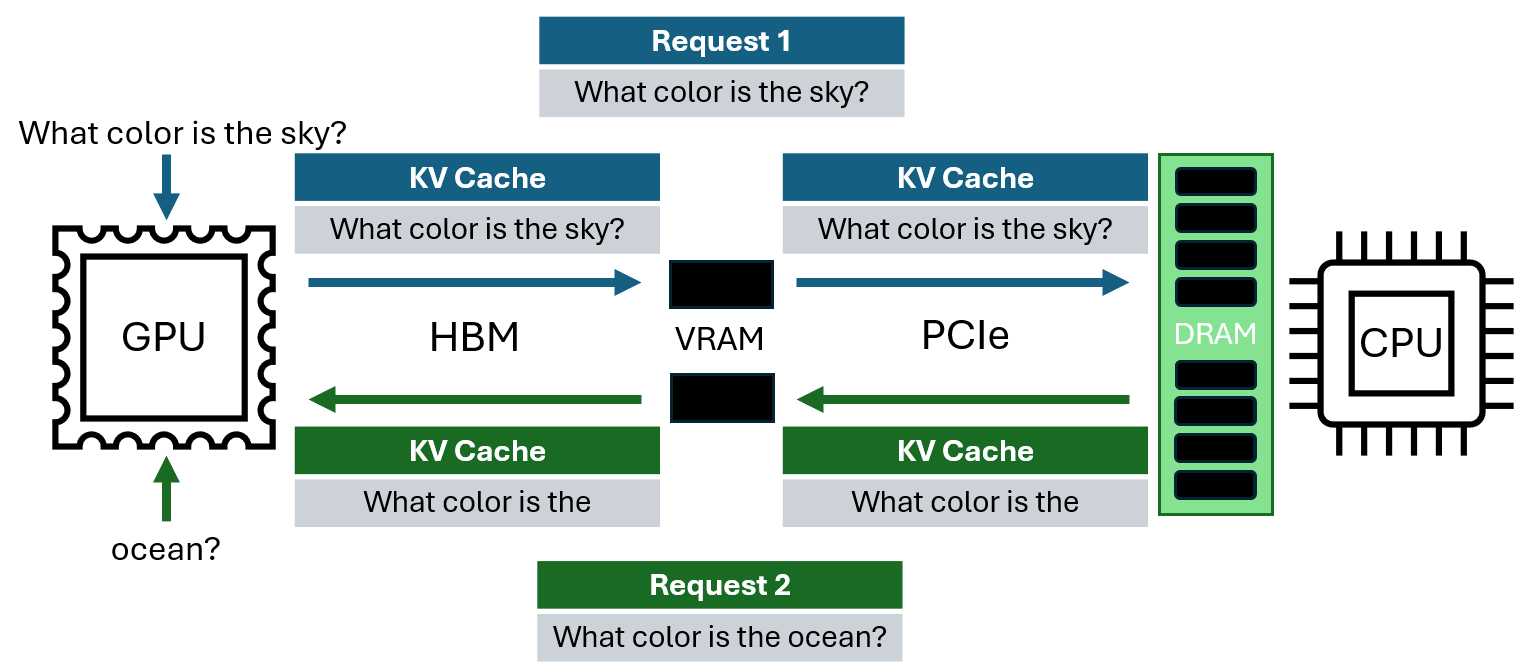}
\caption{KV cache offloading transfers computed representations from GPU VRAM to CPU DRAM. When serving the second request (green), only the novel token ("ocean") requires computation; cached tokens are loaded via PCIe.}
\label{fig:kv_offload}
\end{figure}

KV cache offloading addresses VRAM constraints by placing KV in CPU DRAM, where terabytes of capacity are available (Figure \ref{fig:kv_offload}). However, while GPUs provide terabytes per second of HBM bandwidth, CPU-GPU PCIe interconnects provide only tens of gigabytes per second—orders of magnitude lower. For workloads where cached tokens substantially outnumber new ones, this transforms traditionally compute-bound prefill into memory-bound execution, leaving GPU compute resources underutilized.

Despite growing use of KV offloading, performance implications remain poorly understood. Prior work focuses on optimizing transfer mechanisms \cite{yao2025cacheblendfastlargelanguage} rather than characterizing when and how severely memory bottlenecks constrain performance. This paper develops analytical and empirical frameworks to reveal fundamental trade-offs between computation throughput and memory bandwidth.

We analyze the ratio of offloaded tokens to prefilled tokens, defining $\kappa_{\mathrm{ratio}}$ (workload balance between memory and compute intensity) and $\kappa_{\mathrm{crit}}$ (hardware ratio where computation transitions from compute-bound to memory-bound). This formulation decouples model from system architecture, enabling separate reasoning about software and hardware optimizations.

We apply our framework to ShareGPT, Narrative QA, and FinQA on H100 and B200 systems, measuring end-to-end latency, GPU utilization, and scheduling dynamics. We find severe mismatches: workload $\kappa_{\mathrm{ratio}}$ values greatly exceed system $\kappa_{\mathrm{crit}}$ thresholds, placing most inference in the memory-bound regime. Schedulers cannot exploit compute capacity, and GPUs are underutilized waiting for data transfers. End-to-end latency can spend 99\% of time on PCIe transfers. Our contributions:

\begin{enumerate}[wide, labelwidth=!, labelindent=0pt]

\item \textbf{Bottleneck Analysis.} We determine $\kappa_{\mathrm{crit}}$ ranges from 1 to 76 (most $<$15 before accounting for effective PCIe bandwidth). Workloads exceed this by orders of magnitude: median $\kappa_{\mathrm{ratio}}$ values are 100 for conversations and 5000 for document queries. 90\% of requests process fewer than 133 and 30 new tokens, respectively.

\item \textbf{Performance Characterization.} PCIe overheads reach 86$\times$ prefill computation, accounting for 99\% of execution time. GPUs consume only 28\% of TDP on average, indicating severe underutilization.

\item \textbf{Design Implications.} We identify hardware optimizations (NVLink C2C, unified HBM) that increase $\kappa_{\mathrm{crit}}$ by 5.3$\times$ and 48$\times$, model optimizations (MLA attention) reducing data demand by 5$\times$, and scheduling strategies based on computational intensity.

\end{enumerate}

\section{Re-Visiting Prefill and Decode Characteristics}

This section introduces the technical foundation for understanding KV cache offloading performance. We begin by introducing LLM inference phases and their computational characteristics, then explain how repeated tokens motivate prefix caching, which is increasingly constrained by VRAM capacity. KV cache offloading permits prefix caching for larger contexts but creates memory bottlenecks. Finally, we discuss how KV reuse and offloading impact foundational assumptions in broadly adopted LLM optimizations such as disaggregation and iteration-level scheduling.

\subsection{LLM Inference}

Modern large language models are built on the transformer architecture \cite{vaswani2023attentionneed, grattafiori2024llama3herdmodels, deepseekai2025deepseekr1incentivizingreasoningcapability}, which relies on self-attention to process input sequences. LLM inference consists of two phases, prefill and decode, each with distinct computational and memory characteristics.

\textbf{Transformer Attention and KV Cache.} In self-attention, each input token embedding is transformed into query (Q), key (K) and value (V) vectors through projection matrices $W_Q$, $W_K$, and $W_V$. Attention scores are:
\[Attention(Q,K,V) = softmax(\frac{QK^T}{\sqrt{d_k}})V\]
where $\sqrt{d_k}$ is the key vector dimension.

Transformers cache $K$ and $V$ matrices to avoid recomputing attention, enabling efficient autoregressive generation but incurring large memory overheads. For instance, LLama-3.1-405B generates a 500 KB KV cache per token, reaching tens of gigabytes for long inputs.

\textbf{Prefill and Decode.} During prefill, all input tokens are processed in parallel to generate the initial KV cache and produce the first output token. Prefill is compute-intensive and power-hungry, typically saturating GPU compute resources \cite{10.1145/3620666.3651329}. During decode, the model generates output tokens sequentially, attending to the full KV cache, making decode memory-bound. This paper focuses on prefill, which is traditionally compute-bound but becomes memory-bound when offloading large KV caches.

\subsection{KV Cache Reuse and Prefix Caching}

Prefill's computational costs motivate strategies to avoid redundant computation when tokens repeat across requests. Token repetition is pervasive: multi-turn conversations accumulate context, document Q\&A systems process the same document with different queries, and code completion tools repeatedly analyze project context.

Table~\ref{tab:token_redundancy} illustrates a three-turn conversation where each turn re-processes all previous context. Without caching, this computes KV for 20 tokens; with caching, only 8—a 60\% reduction. For document Q\&A with 100K-token documents, each query reuses the entire context, resulting in thousands of times more cached than new tokens.

\begin{table}
\centering
\scriptsize
\begin{tabular}{@{}llcc@{}}
\toprule
\textbf{Turn} & \textbf{Response} &\textbf{Tokens} & \textbf{Cached / Computed} \\
\midrule
1: "Capital of France?" & "Paris" & 3 & 0 / 3 \\
2: "And population?" & "70 M" & 6 & 4 / 2 \\
3: "What about Germany?" & "Berlin"& 11 & 8 / 3 \\
\midrule
\textbf{Total} & {} & \textbf{20} & \textbf{12 / 8} \\
\bottomrule
\end{tabular}
\vspace{0.5em}
\caption{Token redundancy in multi-turn conversations. Prefix caching reuses KV caches from prior turns, reducing computation by 54\%.}
\label{tab:token_redundancy}
\end{table}

\textbf{Prefix Caching Mechanism.} Prefix caching identifies the longest common prefix between a request's input and previously cached sequences \cite{kwon2023efficientmemorymanagementlarge}. Implementations use hash-based matching (vLLM) or radix trees (SGLang). On a prefix hit, the system loads stored KV tensors from VRAM and computes new ones for only the suffix tokens.

\textbf{VRAM Capacity Constraints.} While prefix caching eliminates redundant computation, storing KV caches in GPU VRAM is limited by capacity. Modern GPUs offer 80–192 GB of VRAM, but most is used for weights and activations, leaving limited space for prefix caches.

Long contexts significantly improve LLM output quality \cite{DBLP:journals/corr/abs-2007-01282, DBLP:journals/corr/abs-2112-04426} but place extraordinary demands on VRAM. Recent models like DeepSeek-R1 and LLaMA-3.1-405B support 128K-token contexts \cite{deepseekai2025deepseekr1incentivizingreasoningcapability, grattafiori2024llama3herdmodels}, requiring up to 66GB of VRAM if cached—most of the H100's 80GB capacity. GPT-5, Gemini, and Grok-4 extend contexts to 400K, 1M, and 4M tokens \cite{GPT-5, Gemini, Grok}.

Serving multiple concurrent, long-context requests quickly exhausts VRAM capacity. Median cached contexts exceed 10K tokens for conversations and 65K for document analysis, making VRAM capacity a critical bottleneck.

\subsection{KV Cache Offloading}

Given VRAM constraints, KV cache offloading stores less frequently accessed KV caches in CPU DRAM or disk \cite{cachegen, yao2025cacheblendfastlargelanguage, gao2024costefficientlargelanguagemodel}. When needed, KV caches are transferred to GPU VRAM via PCIe. Although offloading retains the benefits of KV cache reuse, it encounters bandwidth bottlenecks that transform prefill from compute- to memory-bound execution.

GPUs use a tiered memory hierarchy with order-of-magnitude bandwidth differences. HBM, tightly integrated on a silicon interposer, delivers TB/s, while CPU-GPU PCIe 5.0 provides only 64 GB/s—2\% of HBM's bandwidth. Transferring a 50 GB KV cache takes 15 ms from HBM but 800 ms from CPU DRAM.

For workloads where cached tokens significantly outnumber new ones, this bottleneck is severe. In a document Q\&A scenario with a 65K-token document and 32-token question, Llama-3.1-405B must transfer 33 GB over PCIe (~500ms). On an H100 with 2 PFLOP/s throughput, computing 32 new tokens requires ~26 TFLOPs (~12.8 ms)—39× faster than PCIe transfer. Thus, prefill execution is dominated by memory transfers, leaving GPU compute resources idle.

\subsection{Existing Optimizations and Limits}

The differences between prefill and decode motivate varied optimizations. Two dominant approaches, disaggregation and iteration-level scheduling, assume prefill is compute-bound. We describe these techniques and explain why they provide limited benefits when prefill becomes memory-bound.

\textbf{Disaggregation.} Researchers have proposed separating prefill and decode into independent workloads scheduled on different nodes \cite{patel2024splitwiseefficientgenerativellm, zhong2024distservedisaggregatingprefilldecoding, dynamo}. Prefill is compute-bound; decode is memory-bound. Disaggregation enables precise power allocation (prefill draws 70–100\% peak GPU power; decode uses 20–40\%), hardware matching (compute-intensive prefill benefits from H100s; memory-intensive decode suits cheaper A100s), and optimized tensor parallelism \cite{patel2024splitwiseefficientgenerativellm, dynamo}.

When $K >> T$, prefill converges toward decode—both become bandwidth-limited, undermining hardware specialization benefits. Routing prefill to capable H100s makes less sense if prefill requires computation for few new tokens but requires transferring large KV caches.

\begin{figure}
\centering
\includegraphics[width=.6\linewidth]{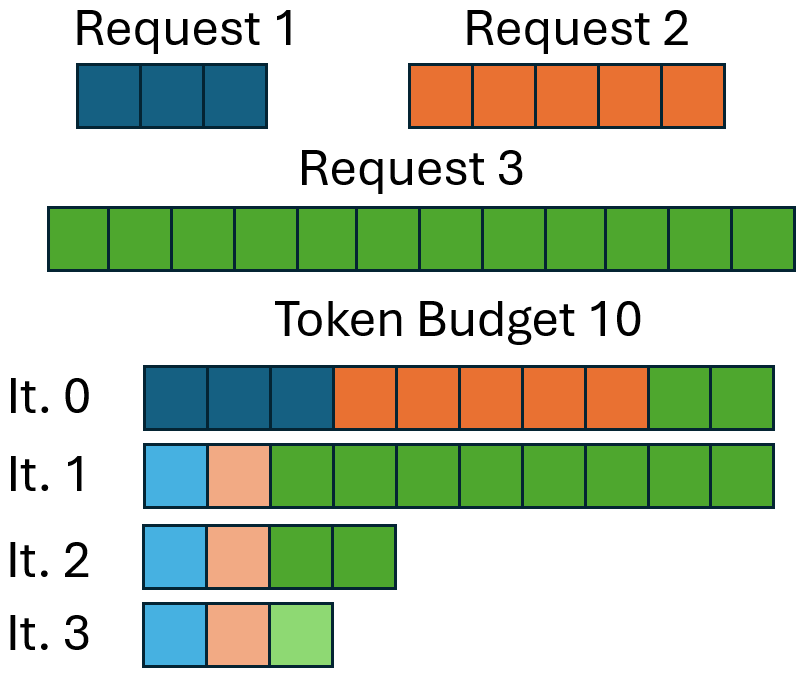}
\caption{Overview of iterative scheduling in the vLLM scheduler \cite{vllm_v1}}
\label{fig:vllm_sched}
\end{figure}

\textbf{Iteration-Level Scheduling.} Iteration-level scheduling improves GPU utilization and reduces TTFT variance \cite{280922} by allocating tokens incrementally. Figure~\ref{fig:vllm_sched} shows vLLM's approach: each request receives a token budget (prefill: input length; decode: one token). The scheduler fills each iteration up to a system-wide budget. The optimal strategy uses the smallest budget that saturates the GPU \cite{dynamo}.

In Fig.~\ref{fig:vllm_sched}, three requests (3, 5, and 12 tokens) share a 10-token budget. Iteration 1 schedules requests 1 and 2 (8 tokens) and partially schedules request 3 (2 tokens). Iteration 2 decodes requests 1 and 2 (2 tokens) and continues request 3's prefill (8 tokens). Batching heterogeneous workloads increases GPU utilization and reduces latency.

vLLM extends iteration-level scheduling for KV cache offloading by deferring offloaded requests until their KV tensors load into VRAM. Token allocation considers only prefill tokens; e.g., a request with 1,000 cached and 100 prefill tokens consumes 100 from the budget.

Scheduling 4K tokens saturates compute on a B200; additional tokens increase TTFT \cite{dynamo}. However, with KV offloading, requests consume memory disproportionately to prefill size, exhausting VRAM before GPU saturation.

\textbf{Emerging Limitations.} Both disaggregation and iteration-level scheduling optimize for compute-bound prefill. KV cache offloading disrupts these assumptions as prefill becomes memory-bound. Disaggregation's hardware specialization becomes ineffective when prefill no longer requires high compute throughput. Iteration-level scheduling's token budgets neglect VRAM requirements for cached contexts, leading requests to exhaust memory before GPUs are fully utilized. The following sections develop an analytical framework to characterize when and how these bottlenecks appear.

\providecommand{\BWpcie}{\mathrm{BW}_{\mathrm{PCIe}}} 

\providecommand{\Ceff}{C_{\mathrm{eff}}}             

\providecommand{\Fpf}{F_{\mathrm{pf}}}               

\providecommand{\Bkv}{B_{\mathrm{kv}}}               

\section{Analytical Framework}
\label{sec:framework}

We develop an analytical framework that determines when KV cache offloading transforms prefill from compute to memory-bound and assesses how this transition impacts performance. The framework accounts for model architecture, hardware specifications, and workload characteristics. 

We describe the system by deriving $\kappa_\text{crit}$, the critical ratio of cached to new tokens at which computation becomes memory-bound. To describe the model and workload, we derive $\kappa_\text{ratio}$, the computation's balance between memory and compute intensity, which depends on the ratio of cached to new tokens. We use these ratios to predict GPU utilization, PCIe overhead, detail how VRAM capacity impacts request concurrency, and detail how schedulers use token budgets. 

\subsection{Memory-Bound Execution}
\label{sec:kcrit}

We model TTFT for a prefill request that loads $K$ cached tokens from CPU DRAM and computes $T$ new prefill tokens on the GPU. Let $F_\text{pf}$ denote the number of FLOPs performed per prefill token, which depends on the model architecture. Let $B_\text{kv}$ denote the byte/token for the model's KV cache, which depends on the number of layers, attention heads, head dimension, and floating-point precision. 
\begin{equation}
B_{\text{kv(GQA)}} = 2 \times L \times H \times d_h \times P \text{ bytes}
\label{eq:kv_size}
\end{equation}

For MLA models, the formula changes to depend on layers, the KV-loRA rank (KVL), and QK-RoPE head dimensions ($d_{QKR}$).
\begin{equation}
B_{\text{kv(MLA)}} = L \times KVL \times d_{QKR} \times P \text{ bytes}
\label{eq:kv_size_mla}
\end{equation}

For system parameters, let $C_\text{eff}$ denote effective GPU FLOPS/s and $\text{BW}_\text{PCIe}$ denote the sustained host-to-device bandwidth in GB/s. Assuming GPU computation and PCIe transfers cannot be fully overlapped, TTFT decomposes into two sequential phases. 
\begin{equation}
\label{eq:ttft}
\text{TTFT} = t_{\text{PCIe}} + t_{\text{prefill}} 
= \frac{K \cdot B_{\text{kv}}}{\text{BW}_{\text{PCIe}}} 
+ \frac{T \cdot F_{\text{pf}}}{C_{\text{eff}}} 
\end{equation}

If transfers and compute partially overlap we can model TTFT as:
\begin{equation}
    \text{TTFT} = (t_{\text{PCIe}}+t_{\text{prefill}}) -\alpha \cdot min(t_{\text{PCIe}},t_{\text{prefill}})
\end{equation}
Where $0<\alpha<1$ depending on the amount of overlap, 1 being perfect overlap and 0 being no overlap. Our measurements indicate that for most representative requests $t_{\text{PCIe}}>>t_{\text{prefill}}$ making equation \ref{eq:ttft} a close approximation regardless of overlap. 
\textbf{Critical Ratio.} Prefill becomes memory-bound when PCIE transfer time dominates GPU compute time. Under these conditions, $t_{\text{PCIe}} > t_{\text{GPU}}$ and 
\begin{align}
K \cdot B_{\text{kv}} &> T \cdot F_{\text{pf}} \cdot \frac{\text{BW}_{\text{PCIe}}}{C_{\text{eff}}} \nonumber \\
\frac{K}{T} &> \frac{F_{\text{pf}}}{B_{\text{kv}}} \cdot \frac{\text{BW}_{\text{PCIe}}}{C_{\text{eff}}} \label{eq:kcrit_derivation}
\end{align}

We define $\kappa_{\text{ratio}} = K/T$ as the ratio of cached to prefill tokens and $\kappa_{\text{crit}}$ as the critical ratio where the transition occurs. Prefill is memory-bound when $\kappa_{\text{ratio}} > \kappa_{\text{crit}}$. $\kappa_{\text{M}}$ and $\kappa_{\text{HW}}$ correspond to the model and hardware factor respectively. 
\begin{equation}
\kappa_{\text{crit}} = \underbrace{\frac{F_{\text{pf}}}{B_{\text{kv}}}}_{\kappa_{\text{M}}} 
\times \underbrace{\frac{\text{BW}_{\text{PCIe}}}{C_{\text{eff}}}}_{\kappa_{\text{HW}}}
\label{eq:kcrit}
\end{equation}

$\kappa_{\text{crit}}$ is dimensionless. In roofline terms, the arithmetic intensity of offloaded prefill is $AI=\frac{T \cdot F_{\text{pf}}}{K \cdot B_{\text{kv}}}$. The compute-to-bandwidth transition occurs where $AI = \frac{C_{\text{eff}}}{\text{BW}_{\text{PCIe}}}$, which rearranges to Equation (5).


\textbf{Model Factors.} The threshold for becoming memory-bound, $\kappa_{\text{crit}}$, increases with computational intensity ($F_{\text{pf}}$) and decreases with KV footprint ($B_{\text{kv}}$). Models with higher $F_{\text{pf}}/B_{\text{kv}}$ ratios, such as those using Multi-Head Latent Attention (MLA) to compress KV representations, exhibit higher $\kappa_{\text{crit}}$ values and are more resilient to memory bottlenecks. 

For example, DeepSeek-V3 is characterized by $B_{\text{kv}} = 70$ KB/token through its use of MLA (Table~\ref{tab:kvtoken}), whereas Qwen3-235B-A22B is characterized by $B_{\text{kv}} = 192$ KB/token for GQA attention. This $2.7\times$ difference means DeepSeek-V3 is less vulnerable to memory bottlenecks. 

\textbf{Hardware Factors.} $\kappa_{\text{crit}}$ increases with 
interconnect bandwidth ($\text{BW}_{\text{PCIe}}$) and decreases with compute throughput ($C_{\text{eff}}$). GPUs that sustain higher throughput are more susceptible to memory bottlenecks as it becomes more difficult to keep the cores fed with data. 

For instance, on an H100 PCIe-5 with 2TFLOP/s, $\kappa_{\text{HW}} \approx 34$; doubling throughput to 4 TFLOP/s halves $\kappa_{\text{HW}}$, requiring twice as many prefill tokens per cached token to stay compute-bound. Conversely, faster interconnects could increase $\kappa_{\text{crit}}$ by an order of magnitude, allowing substantially higher $\kappa_{\text{ratio}}$ workloads to remain compute-bound. More recent interconnect technologies and protocols such as NVLink C2C and PCIe5.0 offer peak bandwidths of 900 GB/s and 64 GB/s respectively. 

\textbf{Case Study with LLaMA-3.1-405B on H100.} We illustrate the calculation for LLaMA-3.1-405B on an H100 GPU with PCIe 5.0. First, examine the model. It performs $F_{\text{pf}} = 2 \times 405 \times 10^9 = 810$ GFLOPs per token, which corresponds to the forward pass through 405B parameters. Table~\ref{tab:kvtoken} indicates the model requires $B_{\text{kv}} = 516$ KB per token. Next, consider the system. PCIe 5.0 peak bandwidth is $\text{BW}_{\text{PCIe}} = 64$ GB/s and the GPU's peak capacity is $C_{\text{eff}} \approx 2000$ TFLOP/s. 

These parameters give $\kappa_{\text{crit}} \approx 50$, so execution becomes memory-bound once cached tokens exceed 50× prefill tokens. A typical 65k token document with 32 prefill tokens yields $\kappa_{\text{ratio}} \approx 2,000$, exceeding $\kappa_{\text{crit}}$ by 40×, firmly placing it in the memory-bound regime. Accounting for measured our measured sustained 15GB/s PCIe bandwidth lowers $\kappa_{\text{crit}} \approx 12$, further confirming memory-bound behavior. Thus, given a model and hardware platform, the architect should determine the system's $\kappa_{\text{crit}}$ and compare against the workload's $\kappa_{\text{ratio}}$ to predict whether requests will be compute or memory-bound. 

\subsection{Resource Utilization}

We define utilization $U$ as the fraction of TTFT spent performing useful computation. From Equation~\ref{eq:ttft}.
\begin{align}
\label{eq:utilization_def}
U 
&= \frac{t_{\text{GPU}}}{\text{TTFT}} 
= \frac{t_{\text{GPU}}}{t_{\text{PCIe}} + t_{\text{GPU}}} 
\end{align}
The PCIe overhead can be thus be defined as:
\begin{align}
\label{eq:poh}
P_{\text{OH}} =\frac{t_{\text{PCIe}}}{t_{\text{GPU}}} = U^{-1} - 1
\end{align}
When VRAM capacity is exhausted by KV cache demands, efficiency can suffer due to under allocation of the scheduler budget. Let $V_{\text{eff}}$ denote the VRAM capacity available for KV caches after allocating space for model weights, activations, and buffers. For requests with 
$K$ cached tokens and $T$ prefill tokens, each request consumes $(K + T) \cdot B_{\text{kv}}$ bytes of VRAM. Equation~\ref{eq:vram_requests} is the maximum number of concurrent requests subject to VRAM constraints and we note that, for requests that are dominated by cached rather than new tokens, $\kappa_{\text{ratio}}$ is high and $K \gg T$. 
\begin{equation}
\label{eq:vram_requests}
N_{\text{max}} = \left\lfloor \frac{V_{\text{eff}}}{(K + T) \cdot B_{\text{kv}}} \right\rfloor
\approx \frac{V_{\text{eff}}}{K \cdot B_{\text{kv}}}
\end{equation}

Each request requires approximately $K \cdot B_{\text{kv}}$ bytes from VRAM capacity but only $T$ tokens from the scheduler's token budget, leading to a significant mismatch between VRAM consumption and scheduler accounting. Equation~\ref{eq:token_budget} is the effective number of prefill tokens processed per iteration, constrained by VRAM capacity. The number of scheduled tokens decreases linearly with $\kappa_{\text{ratio}}$ independent of the token budget. Workloads with high \kratio fully use VRAM capacity well before scheduler capacity. 
\begin{equation}
\label{eq:token_budget}
T_{\text{sched}} = N_{\text{max}} \cdot T 
\approx \frac{V_{\text{eff}}}{K \cdot B_{\text{kv}}} \cdot T
= \frac{V_{\text{eff}}}{\kappa_{\text{ratio}} \cdot B_{\text{kv}}}
\end{equation}

\textbf{Case Study with vLLM on B200.} Consider an B200 deployment with $V_{\text{eff}} = 60$ GB available for KV 
caches, running LLaMA-3.1-405B ($B_{\text{kv}} = 516$ KB/token) with a 4000 token budget. For a document Q\&A request with $K = 65$K cached tokens and $T = 32$ prefill tokens. 
\begin{align}
N_{\text{max}} &= \left\lfloor \frac{60 \text{GB}}{(65,000 + 32) \times 516 \text{KB}} \right\rfloor \approx \lfloor 1.79 \rfloor \text{ requests} \nonumber
\end{align}

The scheduler would only schedule 1.8 requests per iteration, processing $T_{\text{sched}} = 1.8 \times 32 = 57$ 
tokens, $1.4\%$ the iteration's desired 4K . Even for a more moderate workload with $K = 6.4$K cached tokens and $T = 100$ prefill tokens, the scheduler would use only 45\% of the token budget. The 4K token budget was designed for compute-intensive prefill for which token count correlates with computational load. KV cache reuse and offloading breaks this assumption.

\section{Comparative Workload Analysis}
\label{kvcache_quantifying}
\label{sec:workloads}

Section~\ref{sec:framework} derived $\kappa_{\text{crit}}$, the critical ratio 
where prefill transitions from compute-bound to memory-bound. We now measure 
$\kappa_{\text{ratio}} = K/T$ distributions in multi-turn conversations and document question-and-answering to quantify how frequently this transition occurs. 

\subsection{Multi-Turn Conversations}
\label{sec:conversations}

In multi-turn conversations, each turn $i$ processes a new user query ($T_i$ tokens) while attending to all previous exchanges, where $K_i = \sum_{j=1}^{i-1}(Q_j + R_j)$ accumulates queries $Q$ and responses $R$ from prior turns. ShareGPT \cite{sharegpt} contains 90k conversations with 685k turns scraped from ChatGPT. Using LLaMA-3.1 tokenization, user queries average 53 tokens and model responses average 187 tokens. We define $K_i$ as cumulative conversation history and $T_i$ as the current query, assuming prior tokens are offloaded and each turn processes only new tokens \cite{gao2024costefficientlargelanguagemodel}.

\begin{figure}[t]
\centering
\includegraphics[width=\linewidth]{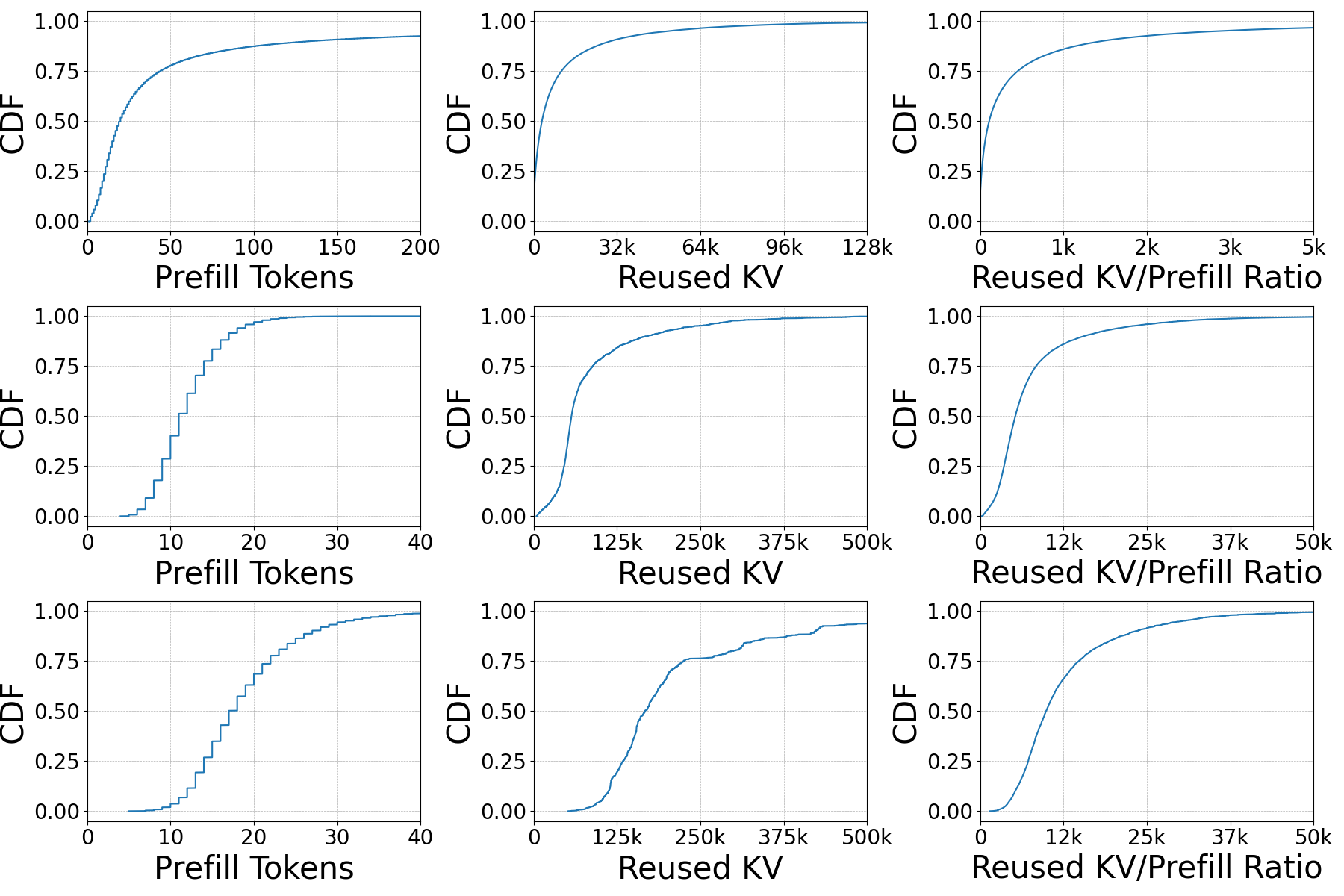}
\caption{Distributions of prefill tokens ($T$), reused KV tokens ($K$), and the ratio of reused KV to prefill tokens ($\kappa_{\text{ratio}}$) for ShareGPT (top), NarrativeQA (middle), and FinQA (bottom). Median Values for $\kappa_{\text{ratio}}$: ShareGPT: 100, NarrativeQA: 5000, FinQA: 10000}
\label{fig:prefill}
\end{figure}

Figure~\ref{fig:prefill} (top) shows multi-turn conversation distributions. Most turns process few tokens: 50\% under 20, 90\% under 133. As turns accumulate, KV caches grow rapidly—50\% exceed 3.2K tokens, 10\% exceed 30K—making cached tokens far larger than prefill tokens. Consequently, 50\% of turns have {\kratio} $>$ 100, and 20\% exceed 1000.

`

\subsection{Document Question Answering}
\label{sec:document_qa}

Document Q\&A queries large documents repeatedly with varied questions. We compute $K$ as document length and $T$ as question length, assuming KV representations for the documents have been computed, cached, and offloaded. NarrativeQA \cite{DBLP:journals/corr/abs-1712-07040} contains 47k questions across 1,572 narrative texts (books, scripts). Documents have median length 57k tokens, ranging from 15K to 190K, and questions are posed with 12 tokens on average. 

FinQA \cite{reddy2024docfinqalongcontextfinancialreasoning} contains 7.4k questions 
across 801 financial documents. Documents have median length of 167k tokens, ranging from 60K to 450K, and questions are posed with 23 tokens on average. 

Figure~\ref{fig:prefill} (middle, bottom) shows $\kappa_{\text{ratio}}$ for querying documents exceeds multi-turn conversations. NarrativeQA has median 5,000 (IQR 2,400–11,000), and FinQA 10,000 (IQR 5,000–17,000), 2 orders of magnitude above ShareGPT’s 100, reflecting much longer documents in Q\&A workloads.

\section{Comparative Platform Analysis}
\label{sec:platform}

Section~\ref{sec:framework} illustrated $\kappa_{\text{crit}}$ for a single platform, LLaMA-3.1-405B on H100 with PCIe 5.0. We now apply the framework to diverse models and hardware systems to assess how platforms impact compute- and memory-boundedness. We examine hardware configurations, showing compute scaling has outpaced bandwidth improvements, and model architectures, comparing dense and sparse MoE as well as grouped-query attention (GQA) and multi-head latent attention (MLA).

\subsection{Hardware Sensitivity}
\label{subsec:hardware_sens}

\begin{table}
\footnotesize
    \centering
    \begin{tabular}{c c c}
    \hline
    \textbf{GPU} & \textbf{PCIe-4} & \textbf{PCIe-5}  \\
    \hline
    B200 & 6.7 & 13.5 \\
    H100 & 17 & 34 \\
    A100 & 53.8 & 107.5 \\
    \hline
    \end{tabular}
\caption{$\kappa_{\text{HW}} = \frac{\text{BW}_{\text{PCIe}}}{C_{\text{eff}}}$ for modern PCIe and GPUs. Lower values indicate greater memory constraints. Values in KB/GFLOP.}
\label{tab:kcritmodel}
\end{table}

Table~\ref{tab:kcritmodel} shows $\kappa_{HW}$ for three GPUs and two PCIe generations; higher values indicate less memory-bound hardware (Eq.~\eqref{eq:kcrit}). Compute improvements have outpaced interconnect bandwidth scaling, making newer GPUs more prone to PCIe bottlenecks. B200 systems exhibit 2.5$\times$ higher $C_{\text{eff}}$ than H100 while PCIe 5.0 provides only 2$\times$ higher bandwidth than PCIe 4.0. This imbalance results in B200 having consistently lower $\kappa_{HW}$ values. B200 with PCIe 5.0 achieves $\kappa_{HW} = 13.5$, only 40\% of H100's 34, meaning B200 transitions to memory-bound execution at 2.5$\times$ lower workload $\kappa_{\text{ratio}}$ values.

\subsection{Model Sensitivity}
\label{subsec:model_sens}

Table~\ref{tab:kvtoken} shows $\kappa_{\text{M}}$ for representative architectures (dense vs. sparse MoE) and attention mechanisms (GQA vs. MLA). We define $\kappa_{\text{M}} = F_{\text{pf}} / B_{\text{kv}}$ using $F_{\text{pf}} = 2N$ FLOPs/token \cite{kaplan2020scalinglawsneurallanguage}, where $N$ is the number of active parameters per forward pass.

\begin{table*}
\footnotesize
    \centering
    \begin{tabular}{l l l l l l l l l}
        \hline
        \textbf{Model} & \makecell{\textbf{Active} \\ \textbf{Params}}  & \textbf{Attention} & \textbf{Arch} & \textbf{$\Bkv$} & $\kappa_{\text{M}}$ & \textbf{$\kappa_{HW}({B200)}$}& \textbf{$\kappa_{HW}({H100)}$} & \textbf{$\kappa_{HW}({A100)}$} \\ \hline
        LLama-3.1-70B & 70B & GQA & Dense & 328 & .42 & 5.7 & 14.3 & 22.6 \\ 
        LLama-3.1-405B & 405B & GQA & Dense & 516 & 1.42 & 19.2 & 48.3 & 76.4\\ 
        Qwen3-30B-A3B & 3.3B & GQA & MOE & 98 & .07 & 1 & 2.4 & 3.8\\
        Qwen3-235B-A22B & 22B & GQA & MOE & 192 & .23 & 3.1 & 7.8 & 12.4\\
        Deepseek-V3 & 37B & MLA & MOE & 70 & 1.06 & 14.3 & 36 & 57\\
        \hline
    \end{tabular}
    \caption{Active parameters, KV/Token ($\Bkv$), $\kappa_{\mathrm{crit}}(\text{model})$ and $\kappa_{\mathrm{crit}}$ for popular models. B200 and H100 use PCIe-5; A100 uses PCIe-4.}
    \label{tab:kvtoken}
\end{table*}

\textbf{Effect of Mixture-of-Experts.} MoE models activate only a fraction of parameters, substantially reducing $F_{\text{pf}}$ compared to dense models. However, $B_{\text{kv}}$ does not scale with active parameters. Qwen3-235B-A22B activates only 5\% as many parameters as LLaMA-3.1-405B, yet its $B_{\text{kv}} = 192$ KB is 37\% of LLaMA-405B's 516 KB. This asymmetry—MoE reduces compute faster than memory footprint—decreases $\kappa_\text{M}$, making MoE models more vulnerable to memory bottlenecks despite computational efficiency.


\textbf{Effect of Multi-Head Latent Attention.} MLA significantly reduces $B_{\text{kv}}$ through low-rank KV projections. DeepSeek-V3 achieves the highest $\kappa_{\text{crit}}$ across all hardware platforms for MoE models, demonstrating that MLA substantially reduces memory constraints compared to GQA. DeepSeek-V3's $B_{\text{kv}} = 70$ KB is only 36\% of Qwen3-235B-A22B's 192 KB and 14\% of LLaMA-405B's 516 KB, directly increasing $\kappa_{\text{M}}$. DeepSeek-V3 reports $\kappa_{\text{crit}} = 1.06$, 4.6$\times$ higher than Qwen3-235B's 0.23 and comparable to LLaMA-405B (1.42) despite fewer active parameters.

Our analysis indicates $\kappa_{\text{crit}}$ values remain relatively low even for optimized configurations. State-of-the-art MoE models like Qwen3-235B achieve $\kappa_{\text{crit}} = 3.1$ on B200, while MLA-optimized DeepSeek-V3 reaches 14.3. All values are modest compared to workloads' $\kappa_{\text{ratio}}$ exceeding 100-1,000 (\S~\ref{sec:workloads}), indicating many real-world requests operate in the memory-bound regime where PCIe bandwidth limits performance.

\subsection{Roofline Analysis}

\begin{figure}
\centering
\begin{subfigure}{.25\textwidth}
  \centering
  \includegraphics[width=\linewidth]{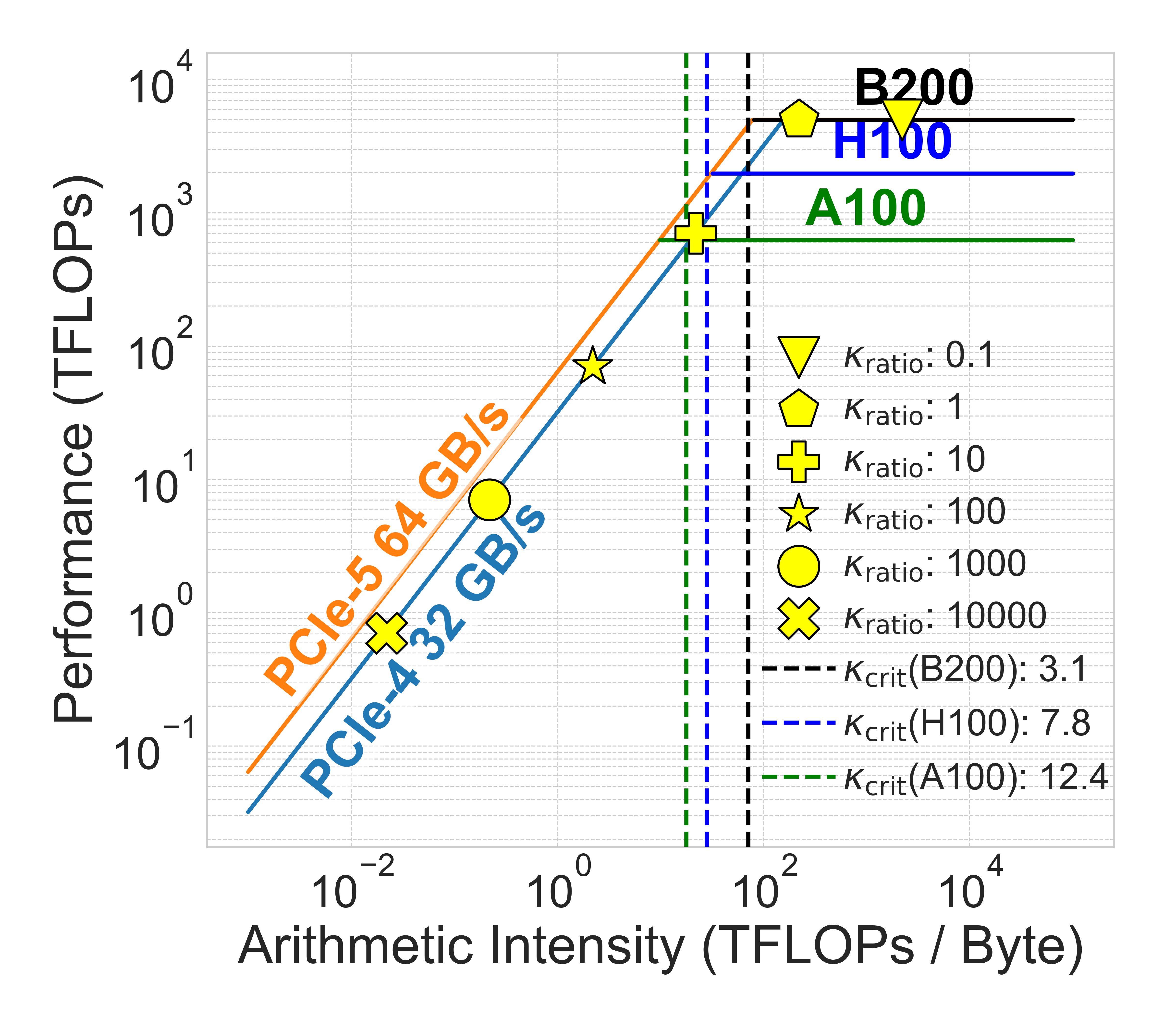}
\end{subfigure}%
\begin{subfigure}{.25\textwidth}
  \centering
  \includegraphics[width=\linewidth]{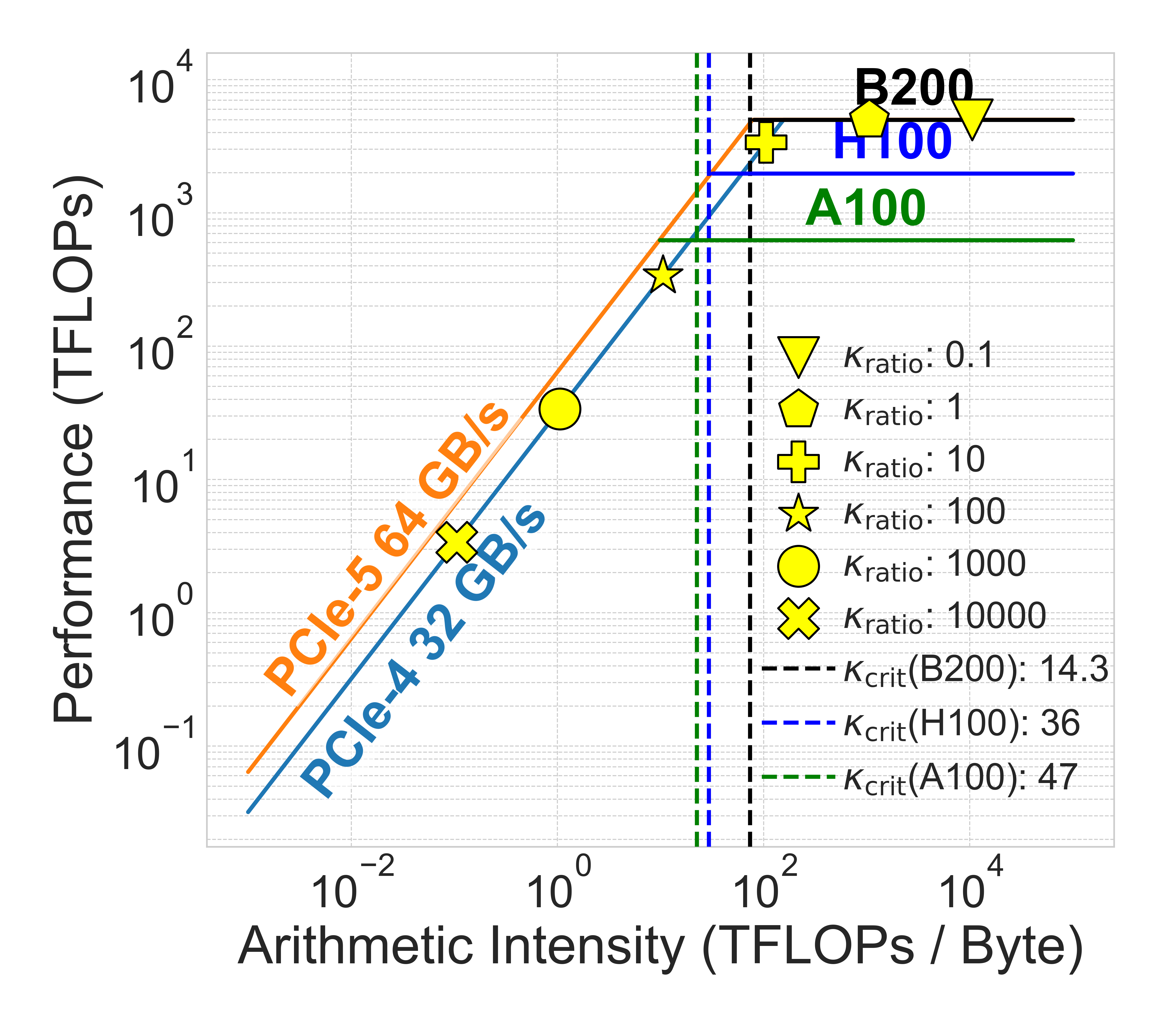}
\end{subfigure}
\vspace{-15pt}
\caption{Roofline models for Qwen3-235B-A22B (left) and Deepseek-V3 (right) on NVIDIA B200, H100 and A100 for current PCIe bandwidths. Vertical lines indicate $\kappa_{\mathrm{crit}}$ for specific hardware. Points left of $\kappa_{\mathrm{crit}}$ are compute-bound; right are bandwidth-bound.}
\label{fig:rooflines}
\end{figure}

Figure~\ref{fig:rooflines} illustrates how architecture and model differences affect performance using roofline analysis \cite{williams2009roofline}. The roofline plots achievable throughput (FLOP/s) against arithmetic intensity (FLOPS/byte), bounded by a horizontal compute ceiling at peak FLOP/s and a diagonal bandwidth ceiling with slope equal to memory bandwidth.

For KV cache offloading, larger $\kappa_{\text{ratio}}$ implies proportionally more data transfer per computed token, reducing FLOPs/byte and pushing the operational point leftward toward bandwidth-limited regime. $\kappa_{\text{crit}}$ marks where the operational point crosses from compute to bandwidth ceiling.

Figure~\ref{fig:rooflines} compares Qwen3-235B-A22B and DeepSeek-V3 across three GPUs and two PCIe generations. At low $\kappa_{\text{ratio}}$, both operate near the compute ceiling. As $\kappa_{\text{ratio}}$ increases past $\kappa_{\text{crit}}$, performance transitions to the bandwidth ceiling, confirming PCIe becomes the bottleneck. DeepSeek-V3's 4.6$\times$ higher $\kappa_{\text{crit}}$ extends the compute-bound region to $\kappa_{\text{ratio}} \approx 40$ versus Qwen3's $\approx$8. However, both become bandwidth-limited at $\kappa_{\text{ratio}} > 100$, typical of real workloads (\S~\ref{sec:workloads}), showing model optimizations alone cannot overcome PCIe bottlenecks.

\textbf{Hardware Effects.} Faster GPUs exacerbate memory bottlenecks under KV offloading. B200 and H100 become memory-bound at lower $\kappa_{\text{ratio}}$ than A100 due to higher peak compute. The B200's 2.5$\times$ higher compute without proportional bandwidth increases exacerbates memory bottlenecks.

\textbf{Model Effects.} MLA compression delays but does not eliminate the memory-bound transition. DeepSeek-V3 maintains compute-intensive execution up to $\kappa_{\text{ratio}} \approx 40$, while Qwen3-235B transitions at $\kappa_{\text{ratio}} \approx 8$. However, both become bandwidth-limited at $\kappa_{\text{ratio}} = 100$-1000, typical of conversation and document Q\&A workloads. Even aggressive KV compression cannot eliminate PCIe bottlenecks.

\section{Experimental Evaluation}

\subsection{Methods}

\textbf{Hardware.} We deploy models on a server with 8$\times$H100 SXM5 GPUs (80GB HBM3 each, connected via NVLink 4.0), an AMD EPYC 7R13 CPU (48 cores), and 2 TB of DDR4-3200 DRAM. GPUs connect to the host via PCIe 5.0 $\times$16 (peak bidirectional bandwidth: 128 GB/s).

\textbf{Software.} We use vLLM v0.10.1~\cite{kwon2023efficientmemorymanagementlarge} with LMCache v0.3.5~\cite{cachegen, yao2025cacheblendfastlargelanguage, LMCache} for KV cache offloading. Models run with FP16 precision. We disable prefix caching to isolate PCIe transfer performance and limit output generation to 1 token to simulate a prefill-only server.

\textbf{Models.} We deploy two models: Llama-3.1-70B (dense transformer with GQA, $\Bkv = 328$ KB/token) and Qwen3-235B-A22B (sparse MoE with GQA, $\Bkv = 192$ KB/token, 22B active of 235B total parameters). We attempted to evaluate DeepSeek-V2 for MLA characterization but encountered implementation-specific overheads preventing accurate PCIe transfer isolation; we defer comprehensive MLA evaluation to future work.

\textbf{Microbenchmarked Measurements.} For each configuration, we issue 200 requests and report mean TTFT with standard deviation using vLLM's benchmark tool. To simulate KV-offloaded requests with $K$ cached and $T$ prefill tokens, we construct prompts as $K$ repetitions of ``Hi'' followed by $T$ randomly sampled single-token words, ensuring exact control while triggering offloading. We measure baseline GPU computation time ($t_{\text{GPU}}$) with offloading disabled and requests with T prefill tokens to isolate PCIe transfer overhead. After a 30-second warm-up, we sample system metrics at 200ms intervals using NVIDIA-smi~\cite{nvsmi} and report average values.

\textbf{Workload Measurements.} We sample 1,000 requests from ShareGPT~\cite{sharegpt} and NarrativeQA based on empirical $\kappa_{\text{ratio}}$ distributions and replay them at 70 and 130 RPS. We instrument vLLM's scheduler to log per-iteration statistics. Each configuration runs for 5 minutes after warm-up to reach steady state.

\subsection{Validating the Analytical Framework}
\label{sec:framework-validation}

\begin{figure}
\centering
\includegraphics[width=0.8\linewidth]{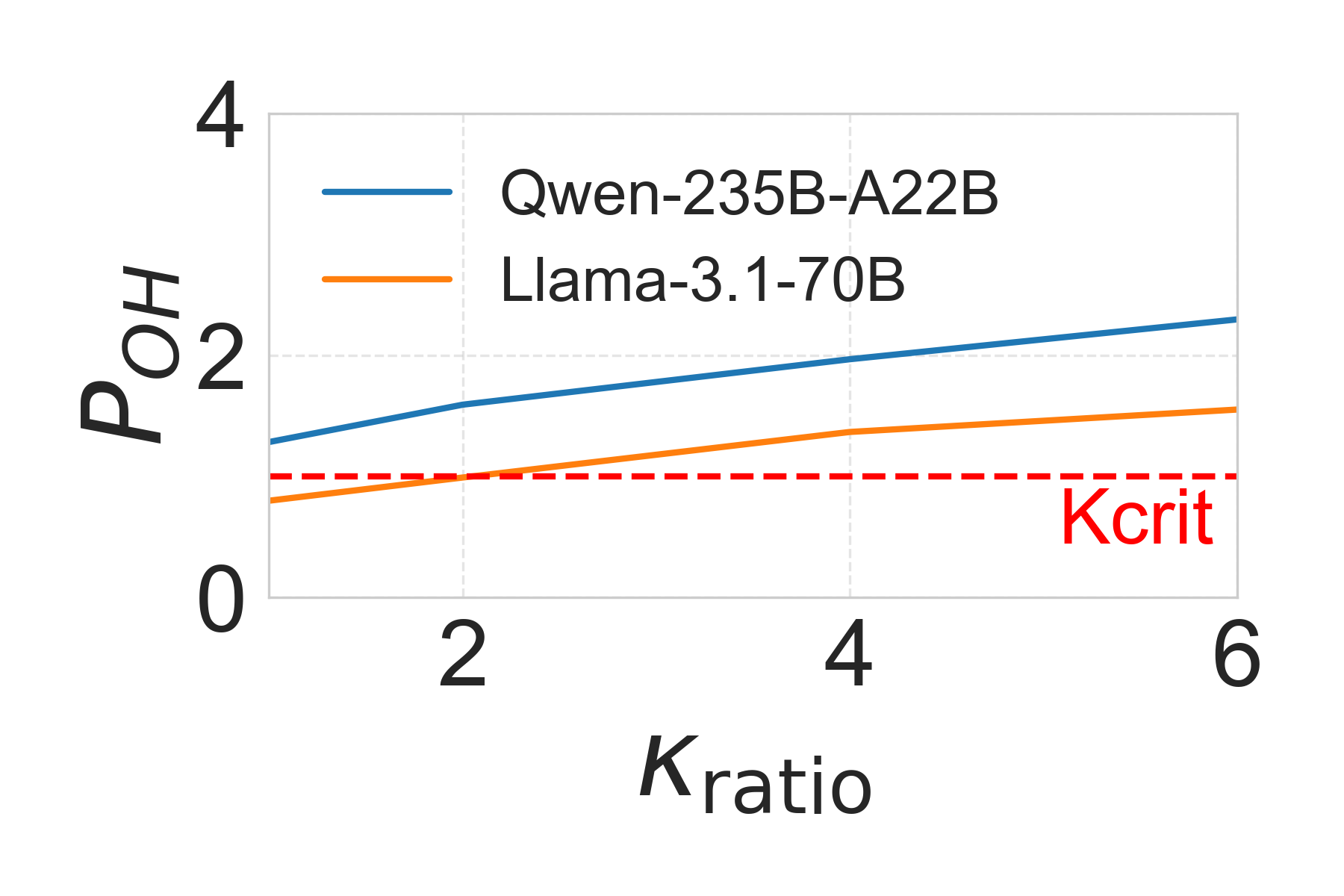}
\vspace{-15pt}
\caption{Measurements of $\kappa_{\text{crit}}$.}
\label{fig:kcrit_eval}
\end{figure}

\textbf{Limits of Analyzing Peak Bandwidth.} Figure~\ref{fig:kcrit_eval} plots PCIe overhead (Eq~\ref{eq:poh}) for varied $\kappa_{\text{ratio}}$ values. Our framework overestimates $\kappa_{\text{crit}}$ (where $P_{OH}=1$ and $t_{\text{PCIe}}=t_{\text{GPU}}$). We measure $\kappa_{\text{crit}}$ values of 2 and 1 versus estimates of 14.3 and 7.8 for Llama and Qwen, respectively. While our model captures the correct functional relationship, it overestimates when using peak hardware specifications. 

KV offloaded workloads become memory bound at extremely small $\kappa_{\text{ratio}}$ values. Over 50\% of execution is spent on PCIe transfers when $\kappa_{\text{ratio}}$ is as low as 1, confirming the memory bottleneck is more severe than hardware specifications suggest.

\textbf{Benefits of Analyzing Actual Bandwidth.} We measure sustained PCIe bandwidth of 15 GB/s (23\% of unidirectional 64 GB/s peak) due to system bottlenecks including CPU-GPU memory copy overheads, NUMA effects, and transfer granularity. Recalculating $\kappa_{\text{crit}}$ using measured bandwidth yields 3.3 and 1.8 for Llama and Qwen respectively, similar to empirical observations. This validates our analytical framework while highlighting the importance of using empirically-derived bandwidth values.

\subsection{Impact of Workload Parameters}

\begin{figure*}
\centering
\begin{subfigure}{.33\textwidth}
  \centering
  \includegraphics[width=\linewidth]{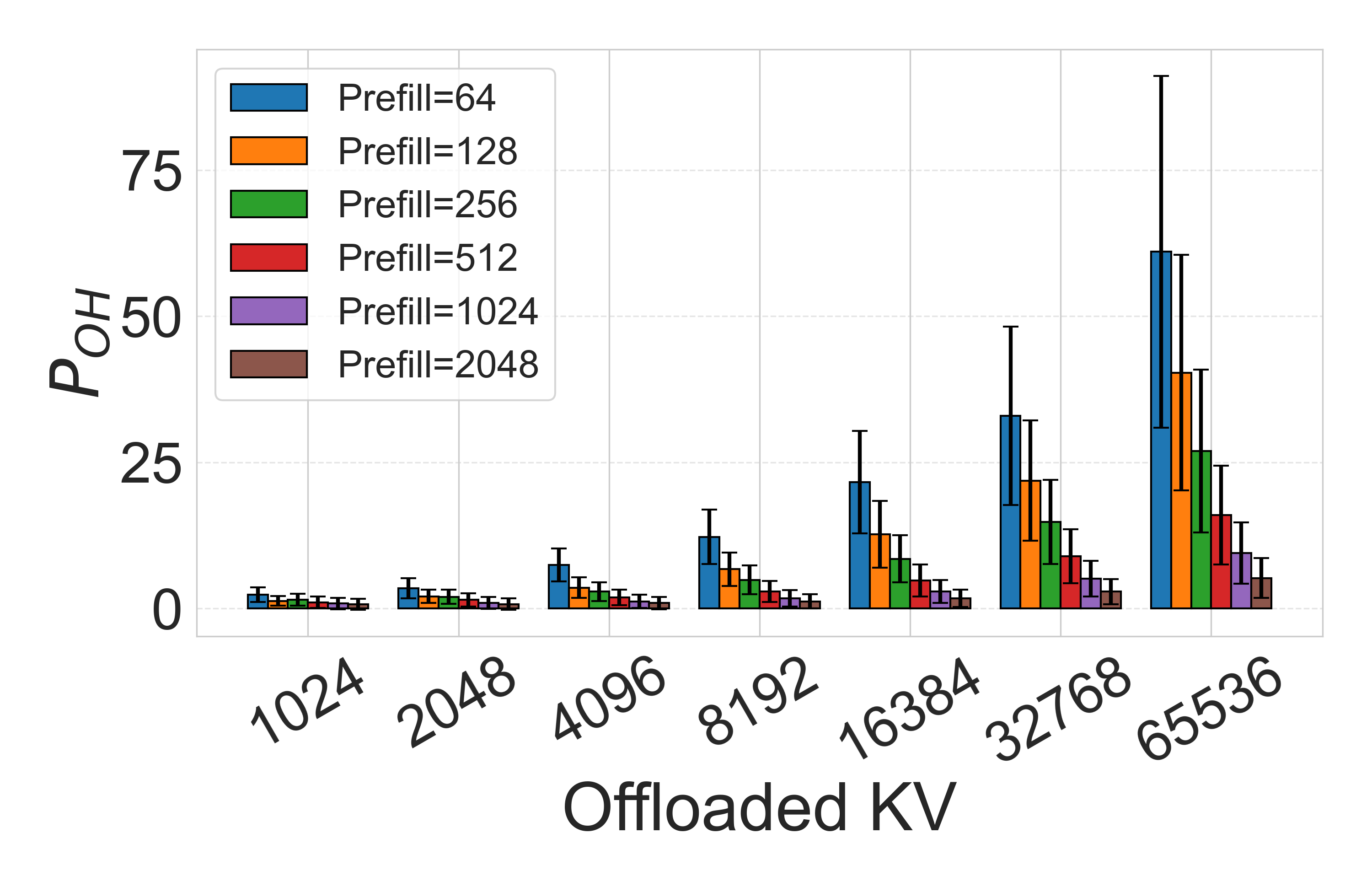}
  \caption{Llama-3.1-70B}
  \label{fig:llama}
\end{subfigure}%
\begin{subfigure}{.33\textwidth}
  \centering
  \includegraphics[width=\linewidth]{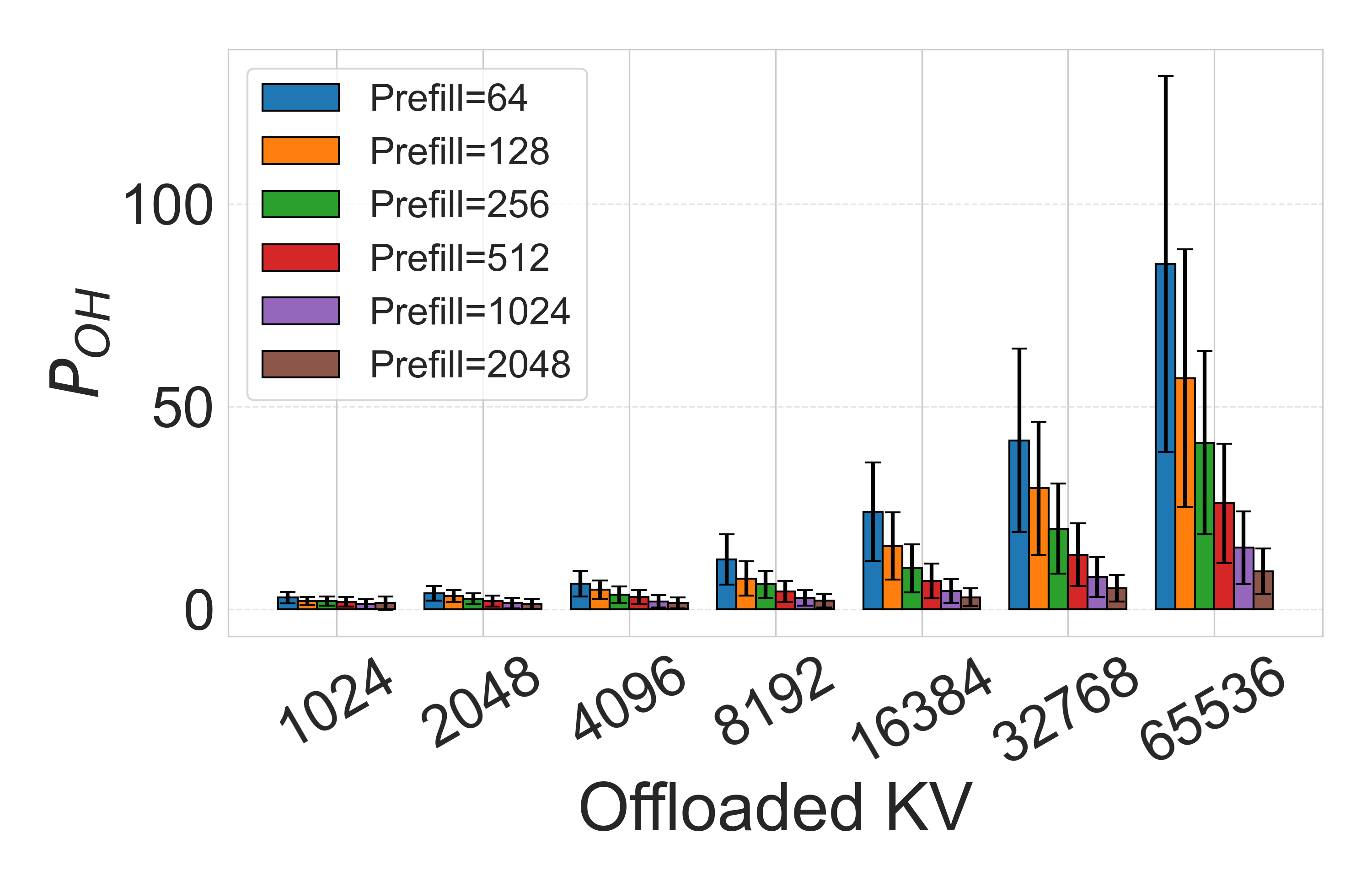}
  \caption{Qwen3-235B-A22B}
  \label{fig:qwen}
\end{subfigure}%
\begin{subfigure}{.33\textwidth}
  \centering
  \includegraphics[width=\linewidth]{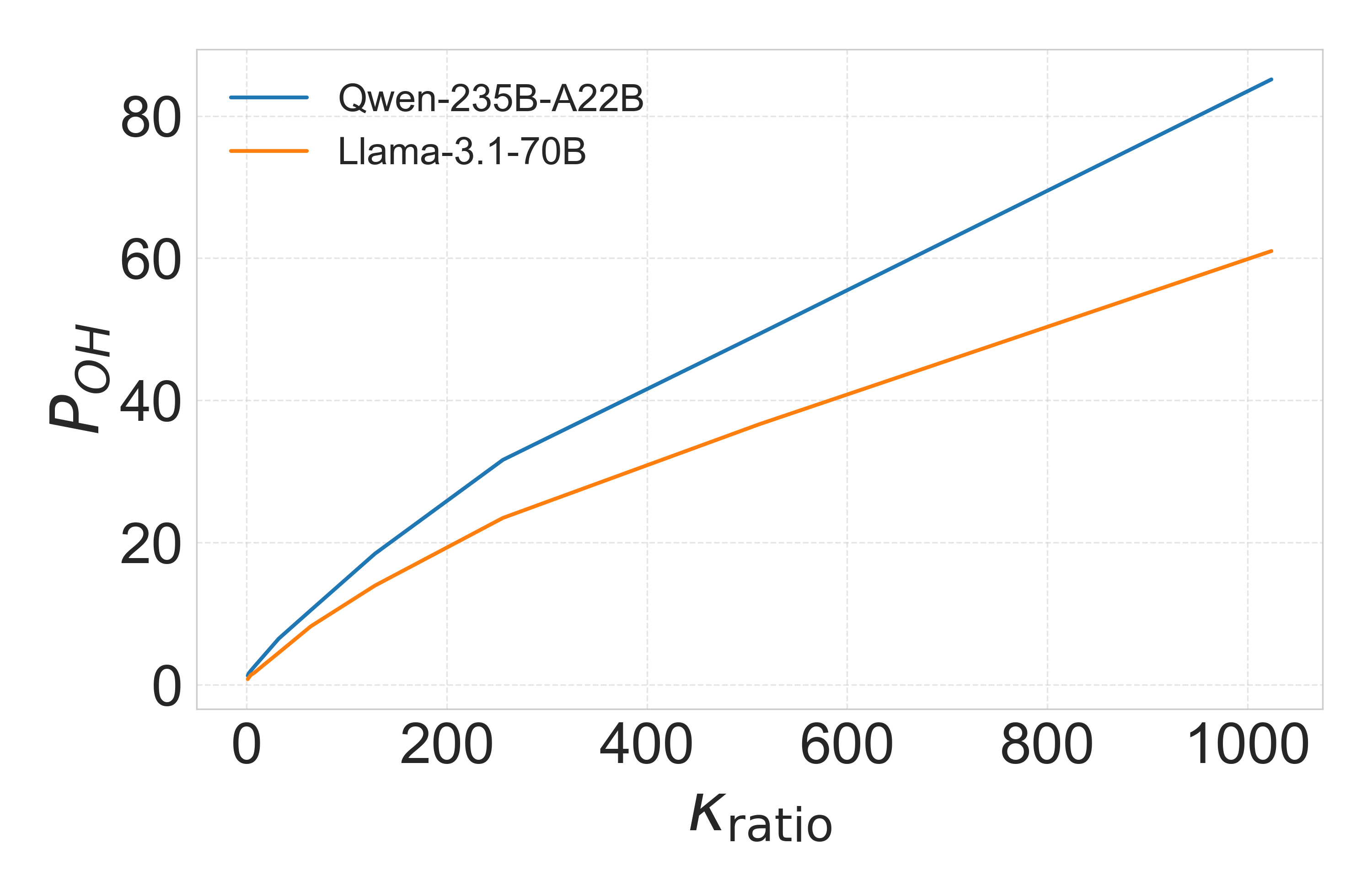}
  \caption{PCIe Overhead vs $\kappa_{\text{ratio}}$}
  \label{fig:poh_vs_ratio}
\end{subfigure}
\caption{PCIe overhead of KV Cache offloading under varying number of prefill tokens and KV cache size. Note, Qwen with $K=65k$, $T=64$ exhibits $P_{\text{OH}}=86$, corresponding to 99\% of execution time spent on PCIe transfers.}
\label{fig:kv_loading}
\end{figure*}

Figure~\ref{fig:kv_loading} characterizes how PCIe overheads vary with workload parameters and model architecture, extending Figure~\ref{fig:kcrit_eval} across the full $\kappa_{\text{ratio}}$ range. Error bars represent TTFT standard deviation from vllm bench serve.

KV-offloaded requests are dominated by PCIe transfers. For a small document with 65K+ KV tokens and 64 input tokens, Qwen spends 99\% of time on transfers. A multi-turn request with 128 inputs and 8K KV spends 88\% on transfers. As $\kappa_{\text{ratio}}$ increases, transfer overheads increase and GPU utilization decreases, aligning with our analytical model.

The Qwen MoE model exhibits consistently higher overheads than the dense Llama model because MoEs reduce computational overheads, exacerbating memory constraints. Thus, computational efficiency from MoEs comes at the cost of increased sensitivity to memory bottlenecks when KV reuse is high.

\subsection{Impact on System Utilization}
\label{sec:system-implications}

\textbf{Scheduler Utilization.} vLLM uses iteration-level scheduling with token budgets to saturate GPU compute. Table~\ref{tab:sgpt_sim} shows average scheduled tokens per iteration for simulated ShareGPT and NarrativeQA requests. For ShareGPT, $K=11115$ and $T=82$ on average. Using Equation~\ref{eq:token_budget} and $V_{\text{eff}}=92$ GB gives $T_{\text{sched}} = 3509$, aligning with measured values. This confirms VRAM capacity, not token budget, limits scheduler throughput. NarrativeQA exhibits much lower $T_{\text{sched}}$ due to increased context length and smaller prompt length.

\begin{table}
\footnotesize
\centering
\caption{Average scheduled tokens per iteration for ShareGPT and NarrativeQA.}
\label{tab:sgpt_sim}
\begin{tabular}{ccc}
\toprule
\textbf{RPS} & \textbf{ShareGPT $T_{\text{sched}}$} & \textbf{NarrativeQA $T_{\text{sched}}$} \\
\midrule
70   &  4064 & 532 \\
130  &  3793 & 555 \\
\bottomrule
\end{tabular}
\end{table}

Schedulers account for tokens in the budget but not KV cache VRAM footprint. A request with 1,000 cached and 100 prefill tokens consumes only 100 tokens from the budget but requires significantly more VRAM. This mismatch causes VRAM to exhaust before the token budget saturates.

\textbf{Power and GPU Utilization.} Table~\ref{tab:sgpt_power} presents average and max power for ShareGPT and NarrativeQA. GPUs use only 28\% and 22\% of maximum TDP respectively. ShareGPT consumes 29\% more power than NarrativeQA due to increased scheduled tokens. Low GPU utilization has significant consequences for datacenters provisioning power infrastructure, cooling, and electrical distribution for peak power. Operating at 22-28\% TDP leaves infrastructure capacity idle, increasing costs per unit of useful work.

\begin{table}[t]
\footnotesize
\centering
\caption{Average and max GPU power for ShareGPT and NarrativeQA.}
\label{tab:sgpt_power}
\begin{tabular}{lcccc}
\toprule
{} &  \multicolumn{2}{c}{\textbf{ShareGPT Power}} & \multicolumn{2}{c}{\textbf{NarrativeQA Power}}\\
\midrule
\textbf{RPS}   &  \textbf{Avg}  & \textbf{Max} &  \textbf{Avg}  & \textbf{Max}  \\
\midrule
70   &  196 & 342   & 152 & 202\\
130   &  192 & 287  & 152 & 194\\
\bottomrule
\end{tabular}
\end{table}

\section{System Implications}


\subsection{Hardware Optimization}
\label{sec:hw-opts}

\begin{figure}
\centering
\begin{subfigure}{.23\textwidth}
  \centering
  \includegraphics[width=\linewidth]{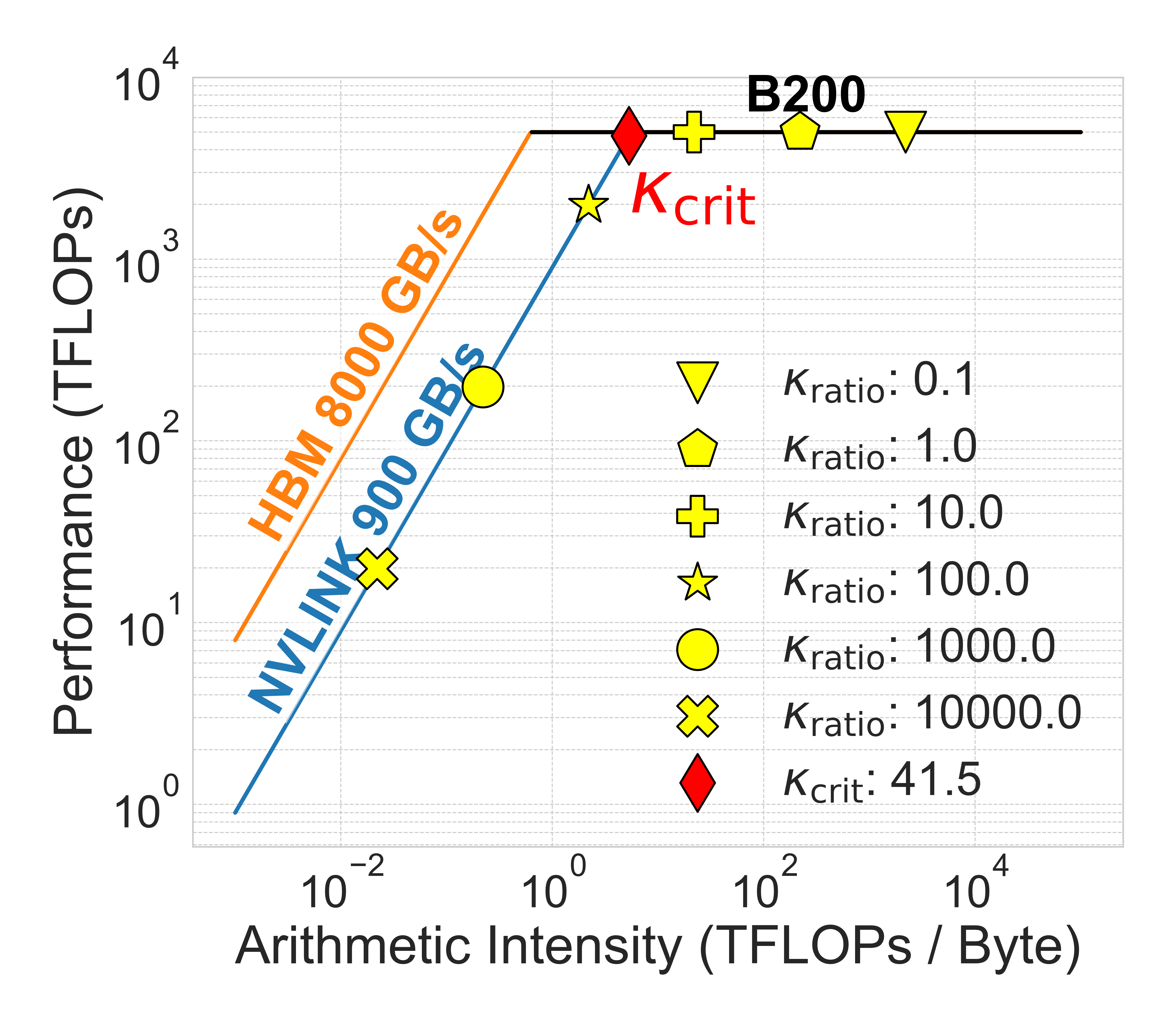}
\end{subfigure}%
\begin{subfigure}{.23\textwidth}
  \centering
  \includegraphics[width=\linewidth]{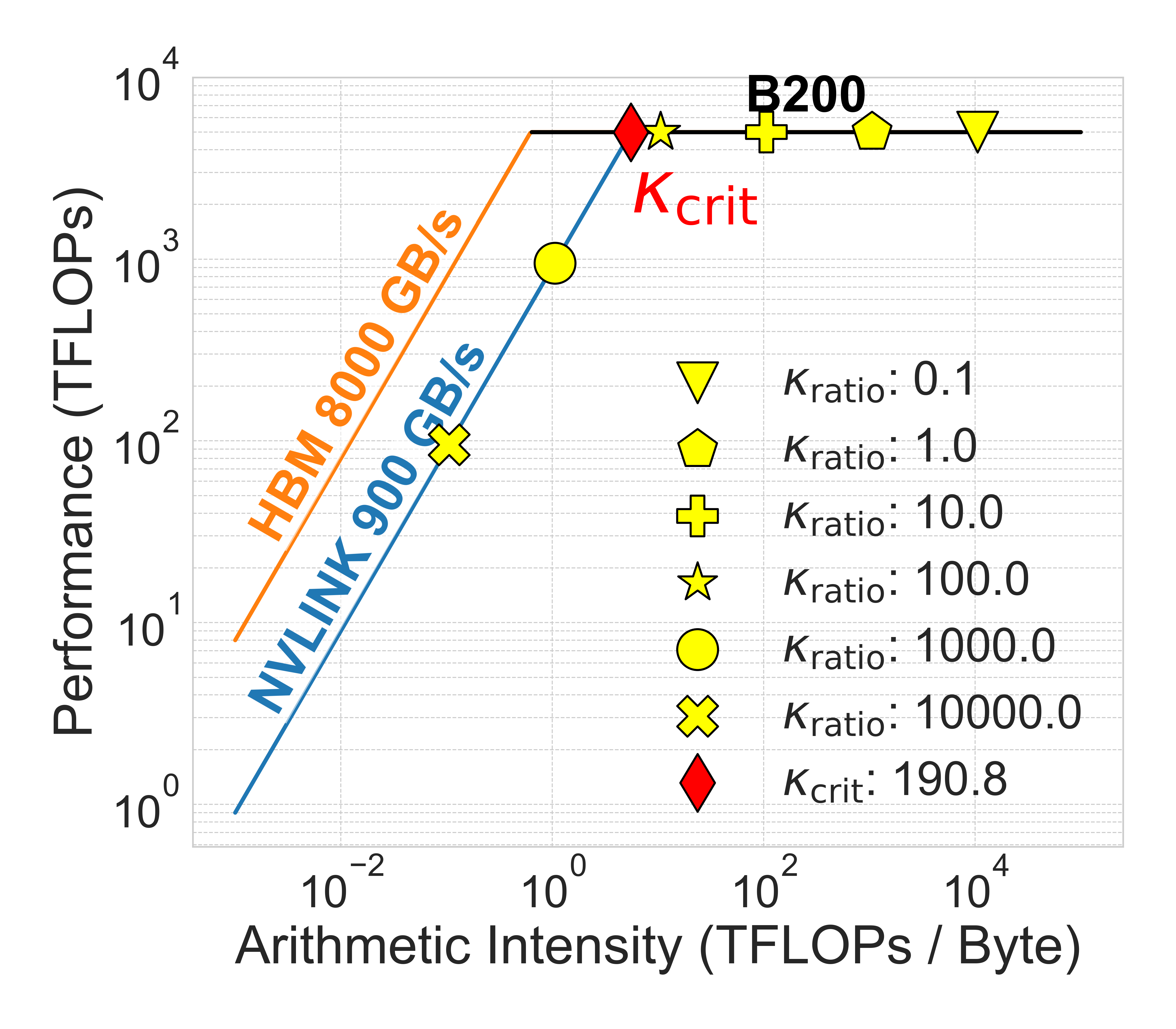}
\end{subfigure}
\caption{Roofline models for Qwen3-235B-A22B (left) and DeepSeek-V3 (right) on NVIDIA GB200 with NVLink C2C and theoretical unified HBM. NVLink C2C increases bandwidth 14$\times$ versus PCIe 5.0, raising $\kappa_{\text{crit}}$ by 5.3$\times$. Unified HBM provides an additional 9$\times$ improvement, dramatically expanding the compute-bound region.}
\label{fig:rooflines_gb}
\end{figure}

\textbf{CPU-GPU Interconnect.} NVIDIA's Grace Blackwell and Grace Hopper employ NVLink Chip-to-Chip (C2C) interconnect with 900 GB/s unidirectional bandwidth, 14$\times$ higher than PCIe 5.0's bandwidth. Figure~\ref{fig:rooflines_gb} presents rooflines for this architecture. Because compute throughput increases only 2.5$\times$ from H100 to B200, $\kappa_{\text{crit}}$ increases by 5.3$\times$. Qwen3-235B-A22B's $\kappa_{\text{crit}}$ improves from 7.8 to 41.5 while DeepSeek-V3's reaches 191. However, many workloads still exceed these thresholds. Document Q\&A's median $\kappa_{\text{ratio}}$ is 5000 and even NVLink C2C systems remain firmly memory-bound. 

\textbf{Unified Memory Architecture.} A more ambitious solution integrates CPU and GPU dies onto the same package with shared access to unified HBM, eliminating the CPU-GPU interconnect bottleneck and allowing both processors to benefit from HBM bandwidth. Such an architecture would provide an order-of-magnitude improvement over NVLink. 

This architecture increases $\kappa_{\text{crit}}$ to 370 for Qwen3-235B-A22B and 1,700 for DeepSeek-V3, a 9$\times$ improvement over NVLink C2C. With such parameters, even document Q\&A workloads approach the compute-bound regime for MLA-based models. While such architectures face challenges in thermal management, die area, and coherency protocols, they represent a promising direction for memory-intensive inference workloads.

\subsection{Model Architecture Optimization}
\label{sec:model-opts}


\textbf{Multi-Head Latent Attention.} MLA, introduced by DeepSeek~\cite{deepseekai2024deepseekv2strongeconomicalefficient}, compresses KV representations through low-rank projections. DeepSeek-V3 achieves $\Bkv = 70$ KB/token versus 192-328 KB/token for GQA-based models (Table~\ref{tab:kvtoken}), a 2.7-4.7$\times$ reduction, that increases $\kappa_{\text{crit}}(\text{model})$ and reduces data transfer overheads proportionally. Realizing MLA's promise requires optimized serving implementations that efficiently handle its compressed KV format. Our evaluation encountered MLA implementation challenges, keeping us from demonstrating these benefits empirically. 

\textbf{KV Cache Quantization.} Quantizing KV caches to lower precision (INT8, FP8, FP4) reduces $\Bkv$~\cite{zhao2024atomlowbitquantizationefficient, yang2024tokenleftbehindreliable}. INT8 quantization provides 2$\times$ reduction with minimal accuracy loss, while aggressive FP4 quantization achieves 4$\times$ compression. Combined with MLA (2.7$\times$), quantization could yield up to 11$\times$ total reduction in memory footprint, substantially shifting workloads back toward compute-bound execution.


\subsection{Workload-Aware Disaggregation}

\textbf{Hardware Routing.} In a heterogeneous cluster with Grace Hopper (NVLink C2C), H100 (PCIe 5.0), and A100 (PCIe 4.0) systems, intelligent routing could match workload characteristics to hardware capabilities. 

\begin{itemize}[leftmargin=*,noitemsep,topsep=2pt]
    \item \textbf{High-$\kappa_{\text{ratio}}$ Requests} ($\kappa_{\text{ratio}} > 100$): Route to Grace Hopper where higher bandwidth reduces transfer overheads.
    \item \textbf{Compute-intensive prefill} ($\kappa_{\text{ratio}} < 1$): Route to H100/A100 for compute throughput at lower cost.
    \item \textbf{Decode requests}: Continue routing to memory-optimized systems as in current disaggregation
\end{itemize}

\textbf{Power Provisioning.} Servers for high-$\kappa_{\text{ratio}}$ workloads can be power-capped to 200-300W without impacting throughput since PCIe bandwidth rather than compute limits performance. This frees power budget for additional servers under the same datacenter power envelope, potentially increasing total throughput through higher concurrency.

Moreover, co-locating memory-intensive requests on the same server quickly exhausts VRAM before saturating compute. Power-aware scheduling should co-locate workloads with complementary characteristics, mixing high-$\kappa_{\text{ratio}}$ and low-$\kappa_{\text{ratio}}$ requests to better utilize VRAM and compute.

\subsection{Utilization-Aware Scheduling}

\begin{figure}[t]
\centering
\includegraphics[width=0.9\linewidth]{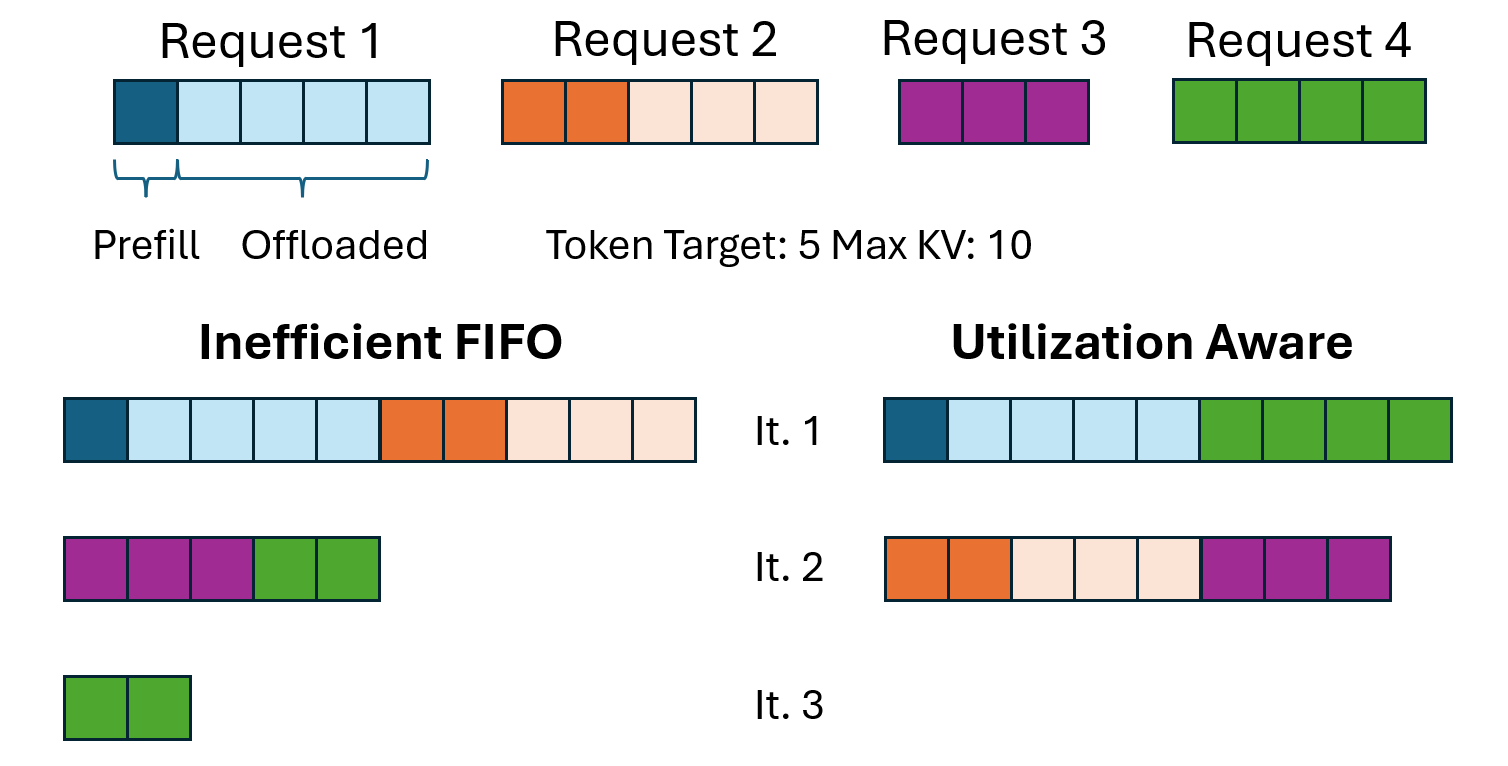}
\caption{Utilization-aware scheduling example. FIFO scheduling (left) underutilizes the token budget (3/5 tokens in iteration 1) due to VRAM constraints. Utilization-aware scheduling (right) reorders requests to saturate the token budget (5/5 tokens) at the cost of delaying earlier requests, improving GPU efficiency while introducing fairness concerns.}
\label{fig:imp_sched}
\end{figure}

\textbf{Scheduling Objective.} Rather than pure token-based accounting, schedulers should maximize compute utilization within VRAM constraints. Figure~\ref{fig:imp_sched} illustrates this approach, describing a system where the optimal token budget per iteration is 5 tokens and the max concurrent tokens due to VRAM constraints is 10. Four requests are queued, each with varying numbers of offloaded and prefill tokens. 

Conventional FIFO scheduling processes requests 1 and 2 in iteration one, using only 3 of 5 available tokens because VRAM capacity is exhausted. An improved scheduler co-schedules requests 1 and 4, utilizing all 5 tokens and reducing total iterations from three to two. This increases GPU efficiency but introduces fairness concerns as request 4 completes before earlier requests 2 and 3.

\textbf{Fairness Mechanisms.} To mitigate starvation of high-$\kappa_{\text{ratio}}$ requests while preserving utilization benefits, schedulers can employ aging credits that increase request priority proportional to wait time. Weighted fair queueing could allocate VRAM proportional to request priority rather than arrival order. And admission control could limit concurrent  high-$\kappa_{\text{ratio}}$ requests to prevent VRAM saturation. 

\section{Related Work}

\textbf{KV Offloading Benchmarking} Previous work \cite{cheng2025lmcacheefficientkvcache, gao2024costefficientlargelanguagemodel} on KV offloading highlight the benefits of KV offloading but ignore its overheads. Our paper addresses this gap by modeling the key bottlenecks and identifying paths for further optimization. Other work \cite{jiang2025kvprefficientllminference} examines PCIe overheads for small, dense models but does not explore how overheads scale with hardware, model, and workload requirements. We extend this to diverse models, hardware, and serving scenarios.

\textbf{Offloading to Other Devices} Other work~\cite{liu2024cachegenkvcachecompression,yao2025cacheblendfastlargelanguage} explores KV offloading to disks or networks. Our framework focuses on CPU DRAM but can be extended to other storage by adjusting $\text{BW}_{\text{PCIe}}$.

\textbf{Extension to Alternative Offloading} Model weights can similarly be offloaded \cite{10.1145/3719330.3721230, liaw2025memascendmemoryoptimizationssdoffloaded, eliseev2023fastinferencemixtureofexpertslanguage, huang2024}. For example, \cite{eliseev2023fastinferencemixtureofexpertslanguage} discusses storing inactive experts in CPU DRAM and only loading active experts. Our framework, which analyzes the relationship between GPU compute and CPU memory, can extend to these cases by adjusting $B_{\text{kv}}$ to describe offloaded expert sizes. 

\textbf{LLM Inference Characterization} Prior LLM inference studies~\cite{10.1145/3620666.3651329} omit modern optimizations like prefix caching and KV offloading. We examine how these techniques reshape computational requirements for prefill. 
\section{Limitations and Future Work}

Our framework approximates $\Fpf$ using $2N$ FLOPs per token, omitting the context scaling term $2 n_\text{layer} n_\text{ctx} d_\text{attn}$~\cite{kaplan2020scalinglawsneurallanguage}. For typical models, this term is negligible. However, for large contexts (e.g., $n_\text{ctx} = 65$K), this term can increase $\Fpf$ by approximately 10\%, introducing minor errors that warrant refinement for very long contexts.

Our framework focuses on loading KV caches from CPU DRAM, ignoring the overhead of storing new prefixes back to DRAM. We assume write operations occur post-prefill and that bidirectional PCIe mitigates impact. However, larger $T$ values may introduce unmodeled overheads from read-write contention.

While our framework shows attention optimizations (MLA, compression) reduce PCIe overheads, we do not empirically characterize these techniques. Future work should measure MLA (2.7-4.7$\times$ $\Bkv$ reduction) and quantization (2-4$\times$ compression) across diverse models and hardware to validate predicted benefits.

\section{Conclusion}

We develop an analytical framework to showcase how prefill becomes memory constrained for various model characteristics and hardware specifications. We empirically characterize KV offloading to see how our model estimates align with real data. We find that prefill becomes memory constrained for very small $K_\text{ratio}$ values resulting in resource waste and GPU underutilization. Based on our findings, we discuss possible directions for hardware development, model architecture changes, and framework adjustments to optimize the capabilities of KV cache offloading.

\nocite{langley00}

\bibliography{example_paper}
\bibliographystyle{mlsys2025}



\end{document}